\documentclass[prd,
,superscriptaddress
,nofootinbib,%
tightenlines ]{revtex4}
\usepackage{epsfig}
\usepackage{amssymb,color}
\usepackage{amsfonts}
\usepackage{amssymb}
\usepackage{slashed}

\newcommand{\ben}{\begin{displaymath}}
\newcommand{\een}{\end{displaymath}}
\newcommand{\be}{\begin{equation}}
\newcommand{\ee}{\end{equation}}
\newcommand{\bea}{\begin{eqnarray}}
\newcommand{\eea}{\end{eqnarray}}
\begin{document}
\title{ Vacuum energy in effective field theory of general relativity 
}

\author{E.~Epelbaum}
\affiliation{Ruhr-University Bochum, Faculty of Physics and Astronomy,
Institute for Theoretical Physics II, D-44870 Bochum, Germany}
\author{J.~Gegelia}
\affiliation{Ruhr-University Bochum, Faculty of Physics and Astronomy,
Institute for Theoretical Physics II, D-44870 Bochum, Germany}
\affiliation{Tbilisi State  University,  0186 Tbilisi, Georgia}
\author{Ulf-G.~Mei{\ss}ner}
 \affiliation{Helmholtz Institut f\"ur Strahlen- und Kernphysik and Bethe
   Center for Theoretical Physics, Universit\"at Bonn, D-53115 Bonn, Germany}
 \affiliation{Institute for Advanced Simulation (IAS-4) , Forschungszentrum J\"ulich, D-52425 J\"ulich,
Germany}
\affiliation{Peng Huanwu Collaborative Center for Research and Education, International Institute for Interdisciplinary and Frontiers, Beihang University, Beijing 100191, China}

\begin{abstract}

It is shown to all orders of perturbation theory that in the effective
field theory of general relativity the condition of vanishing of the
vacuum energy leads to the same value of the cosmological constant,
viewed as a parameter of the effective Lagrangian, as obtained by demanding the 
consistency of the effective field theory in Minkowski background. 
The resulting effective action is characterized by the cosmological
constant term that vanishes exactly. 


\end{abstract}

\maketitle

It is difficult to disentangle the philosophical and physical aspects of the cosmological constant problem. 
Eventually, a lot depends on the attitude one takes towards the relationship between the science and the material world around us.  
It seems natural to think that the vacuum state of our Universe is
what remains if the whole material
is removed, that is, vacuum means just "nothing". 
It seems even more natural that an adequate theory
should assign zero energy to "nothing"
\cite{Faddeev:1982id}. 
Does this "naive" intuitive picture contradict our well-established
and very successful quantum theory? Note that arguments in favor of the non-vanishing vacuum energy 
based on the Casimir effect should, at least, be considered as questionable, see, e.g., Refs.~\cite{Jaffe:2005vp,Nikolic:2016kkp}.
The experimental observation of the accelerating expansion of the universe
(see, e.g., Ref.~\cite{Rubin:2016iqe} and references therein) is often
being interpreted as evidence of the vacuum energy to 
have a nonzero value (for a review see, e.g., Ref.~\cite{Martin:2012bt}).
This leads to the well-known cosmological constant problem, caused by a
huge mismatch  between the theoretical estimations of  the cosmological constant and its value 
suggested by the experimental data  \cite{Weinberg:1988cp}.
This problem is also relevant for modern cosmology, which relies on Einstein's theory of general
relativity.

\medskip
In modern view, Einstein's theory of general relativity is
regarded as a leading-order approximation to an effective field theory (EFT).
It is widely accepted that at low energies, all fundamental interactions including gravity can be described by an EFT \cite{Weinberg:1995mt}. 
 Not every background of general relativity represents a valid vacuum in quantum theory. Various considerations suggest that among the three maximally-symmetric backgrounds, Minkowski is the only consistent vacuum with the possibility of non-trivial cosmological history \cite{Dvali:2013eja,Dvali:2014gua,Dvali:2017eba,Dvali:2020etd,Dvali:2024dlb}. 
In the EFT framework, the cosmological constant is just one of the parameters of the effective Lagrangian of interacting gravitational and matter fields \cite{Donoghue:1994dn,Donoghue:2015hwa,Donoghue:2017pgk}. 
We seem to be rather far from constructing/discovering the fundamental theory underlying this effective theory (if it exists at all), however, we can impose conditions on parameters 
of the low-energy EFT such that the vacuum energy is exactly
zero. Even if such an underlying fundamental theory does not
  exist, the property of  
vanishing vacuum energy can equally well  be realized order-by-order in perturbation theory within the low-energy EFT.
In the realm of perturbation theory, this uniquely fixes the cosmological constant as a function of other parameters of the effective Lagrangian.
On the other hand, the cosmological constant gets fixed from the
consistency condition for the considered EFT in  Minkowski background. 
In particular, the consistency of an EFT in Minkoswski background,
enforced by demanding the absence of the massive ghost graviton
degrees of freedom, was
shown to uniquely determine  the cosmological constant to all orders
in the loop expansion for the case of an Abelian gauge 
theory with spontaneous symmetry breaking coupled to the metric field \cite{Burns:2014bva}. 
At the two-loop level, the conditions of vanishing vacuum energy and the absence of massive
ghost modes were found to yield the same expressions for the
cosmological constant, thanks to 
non-trivial cancellations between different contributions \cite{Gegelia:2019fjx,Gegelia:2019zrv,Epelbaum:2024qdx}. 
Notice that while the authors of Ref.~\cite{Burns:2014bva} consider a
particular Abelian model for demonstration, their derivation of the Ward identity and the graviton low-energy theorem, used to obtain the condition of self-consistency, applies to any local quantum field theory which 
is invariant under transformations of diffeomorphism, including the Standard Model (SM) coupled to general relativity.  
Below we give a general argument imposing that the two-loop-order
results of
Refs.~\cite{Gegelia:2019fjx,Gegelia:2019zrv,Epelbaum:2024qdx} hold  to
all orders of perturbation theory in the EFT of general relativity. That
is, the consistency of the EFT in Minkowski background and the requirement
of the vanishing vacuum energy lead to the same value of the cosmological constant as a parameter of the effective Lagrangian.

\medskip

We consider the action specified by the most general effective Lagrangian of gravitational and matter fields of the SM, which is invariant under 
general coordinate transformations and other underlying symmetries 
\begin{eqnarray}
S = 
\int d^4x \, \sqrt{-g}\, \left\{ \frac{2}{\kappa^2} (R-2\Lambda)+{\cal L}_{\rm gr,ho}
+{\cal L}_{\rm m}\right\},
\label{action}
\end{eqnarray}
where $\kappa^2=32 \pi G$, $G$ is the Newton's gravitational constant,
$g$ denotes the determinant of the metric field $g^{\mu\nu}$,
$\Lambda$ is the cosmological constant and $R$ refers to the scalar
curvature. Further,  ${\cal L}_{\rm gr,ho}$ represents 
self-interaction terms of the gravitational field with higher numbers of derivatives 
while ${\cal L}_{\rm matter}$ is the effective Lagrangian of the matter fields interacting with gravity. 
The success of the theory of gravitation based on the Einstein-Hilbert action suggests that the contributions of 
${\cal L}_{\rm gr,ho}$ are heavily suppressed by some large scale(s).

\medskip

The energy-momentum tensor $T^{\mu\nu}_{\rm m}$ of the matter fields
coupled to the gravitational field and the pseudotensor $T^{\mu\nu}_{\rm gr}$
of the gravitational field are given by 
\begin{eqnarray}
T^{\mu\nu}_{\rm m}   =  \frac{2}{\sqrt{-g}}\frac{\delta S_{\rm m} }{\delta g_{\mu\nu}}\,,
\qquad
T^{\mu\nu}_{\rm gr} = 
\frac{4}{\kappa^2} \, \Lambda\,g^{\mu\nu}+  T_{\rm LL}^{\mu\nu}+\frac{2}{\sqrt{-g}}\frac{\delta S_{\rm gr,ho} }{\delta g_{\mu\nu}}  \,,
\label{defTs}
\end{eqnarray}
\noindent
where  $T_{\rm LL}^{\mu\nu}$ is defined via \cite{Landau:1982dva}
\begin{eqnarray}
(-g)T^{\mu\nu}_{LL}  &=& \frac{2}{\kappa^2} \left(\frac{1}{8} \, g^{\lambda \sigma } g^{\mu \nu } g_{\alpha \gamma}
g_{\beta\delta} \, \mathfrak{g}^{\alpha \gamma},_{\sigma }
   \, \mathfrak{g}^{\beta \delta},_\lambda -\frac{1}{4} \, g^{\mu \lambda } g^{\nu\sigma } g_{\alpha \gamma} g_{\beta \delta }
  \, \mathfrak{g}^{\alpha\gamma},_\sigma \, \mathfrak{g}^{\beta\delta},_\lambda -\frac{1}{4} \, g^{\lambda \sigma } g^{\mu \nu } 
   g_{\beta \alpha} g_{\gamma \delta} \, \mathfrak{g}^{\alpha \gamma},_\sigma \, \mathfrak{g}^{\beta \delta},_\lambda \right.\nonumber\\
&+& \left. \frac{1}{2}\,  g^{\mu \lambda } g^{\nu\sigma } g_{\beta \alpha} g_{\gamma \delta } \, \mathfrak{g}^{\alpha \gamma},_\sigma \, \mathfrak{g}^{\beta\delta},_\lambda 
+g^{\beta \alpha } g_{\lambda \sigma }
  \, \mathfrak{g}^{\nu \sigma},_\alpha \, \mathfrak{g}^{ \mu\lambda},_\beta +\frac{1}{2} \, g^{\mu \nu } g_{\lambda \sigma }
   \, \mathfrak{g}^{\lambda \beta},_\alpha \, \mathfrak{g}^{\alpha\sigma},_\beta \right.\nonumber\\
&-& \left.g^{\mu \lambda } g_{\sigma \beta }
   \, \mathfrak{g}^{\nu \beta},_\alpha \, \mathfrak{g}^{\sigma\alpha},_\lambda 
   -g^{\nu \lambda } g_{\sigma \beta} \, \mathfrak{g}^{\mu\beta},_\alpha \, \mathfrak{g}^{\sigma \alpha},_\lambda
   +\, \mathfrak{g}^{\lambda \sigma},_\sigma \, \mathfrak{g}^{\mu\nu},_\lambda 
   - \, \mathfrak{g}^{\mu \lambda},_\lambda \, \mathfrak{g}^{\nu \sigma},_\sigma \right),
   \label{LLEMT}
\end{eqnarray}
with $\mathfrak{g}^{\mu\nu}=\sqrt{-g} \, g^{\mu\nu}$ and $\mathfrak{g}^{\mu\nu},_\lambda=\partial\mathfrak{g}^{\mu\nu}/\partial x^\lambda $.

The full energy-momentum tensor $T^{\mu\nu}=T^{\mu\nu}_{\rm m}
+T^{\mu\nu}_{\rm gr} $ gives rise to the conserved
 four-momentum of the matter and gravitational fields given by \cite{Landau:1982dva}
\begin{equation}
P^\mu= \int (-g) \, T^{\mu\nu} d S_\nu\,,
\label{EMV}
\end{equation}
where the integration is carried out over any hyper-surface containing the whole three-dimensional  space. 

\medskip

The cosmological constant $\Lambda$ can be uniquely fixed by imposing the condition that the energy of the vacuum is zero. 
This is achieved by demanding that the vacuum expectation value of the integrand in Eq.~(\ref{EMV}) vanishes order-by-order in perturbation theory. 
To show that this condition is equivalent to the self-consistency condition of Ref.~\cite{Burns:2014bva}, consider the vacuum-to-vacuum transition amplitude
\begin{equation}
{\cal I}=\int {\cal D}h {\cal D}\psi {\cal D} \bar c {\cal D} c \, e^{\frac{i}{\hbar} S_E(g,\psi,\bar c,c)}\,,
\label{trans}
\end{equation}
where $\psi$ and $c$, $\bar c$ represent the matter and ghost fields,
respectively, and the metric field is decomposed as
$g^{\mu\nu}=\eta^{\mu\nu}+\kappa h^{\mu\nu}$, with $\eta^{\mu\nu}$ denoting the
the Minkowski  background. Here, the action $S_E$ contains the gauge
fixing and ghost terms in addition to the action $S$ specified in
Eq.~(\ref{action}). Note that the vielbein tetrad formalism has to be used for fermionic degrees of freedom, but we suppress these details here.
This integral remains unaffected by the change of the integration variables 
\begin{equation}
h^{\mu\nu}(x)\to h^{\mu\nu}(x)+\epsilon^{\mu\nu}(x) f(g) 
\label{ChOfVars}
\end{equation}
where $\epsilon^{\mu\nu}(x)$ are the infinitesimal local transformation
parameters and $f(g)$ is an arbitrary non-singular function of $g$. 


Taking into account the Jacobian of the change of variables, 
\begin{eqnarray}
{\cal J} 
= {\rm Tr} \left\{  \left[ 1-  \epsilon^{\alpha\beta}(x) g_{\alpha\beta}(x)  g(x) f'(g(x)) \right]  \delta^4(x-y)  \right\}  
+ {\cal O}(\epsilon^2) \,,
\label{Jakobiantrans}
\end{eqnarray}
we obtain
\begin{eqnarray}
0=\delta {\cal I}  =  
- \int \! d^4y \, \epsilon^{\alpha\beta} (y) \int {\cal D}h {\cal D}\psi {\cal D} \bar c {\cal D} c 
\left( 
  g f'(g) g_{\alpha\beta}(y) \delta^4(0) 
 +\frac{i}{\hbar}\, f(g) \frac{\delta S_E}{\delta g^{\alpha\beta}(y)}\right) e^{\frac{i}{\hbar}\,S_E(g,\psi,\bar c,c)} + {\cal O}(\epsilon^2)\,.
\label{trans}
\end{eqnarray}
Here, the variation of the action is given by  
\begin{equation}
\frac{\delta S_E}{\delta g^{\alpha\beta}(y)} = \frac{\sqrt{-g}}{2}  \left[ T^{\mu\nu}_{\rm m} - \frac{4}{\kappa^2} \left( R^{\mu\nu}- \frac{1}{2} \, g^{\mu\nu} R + \Lambda \, g^{\mu\nu} \right) 
+ {\cal T}^{\mu\nu} \right] \,,
\label{EEOM}
\end{equation}
where ${\cal T}^{\mu\nu}$ denotes the contributions of the gauge fixing and ghost terms, as well as of the higher-order corrections.

\medskip
\medskip
\noindent
Next, using
$ T^{\mu\nu} = T^{\mu\nu}_{\rm LL} +T^{\mu\nu}_{\rm m} - 4 \Lambda \, g^{\mu\nu} /(\kappa^2) + {\cal T}^{\mu\nu} $
and the identity \cite{Landau:1982dva}
\begin{equation}
(-g)\left\{ \frac{4}{\kappa^2}\left( R^{\mu\nu}- \frac{1}{2} \, g^{\mu\nu} R\right) + T^{\mu\nu}_{\rm LL}\right\} = \frac{\partial h^{\mu\nu\lambda}}{\partial x^\lambda} \,,
\label{def1}
\end{equation}
where 
\begin{equation}
 h^{\mu\nu\lambda}  =  \frac{2}{\kappa^2} \frac{\partial}{\partial x^\sigma} \left\{ (-g) \left( g^{\mu\nu} g^{\lambda\sigma} - g^{\mu\lambda} g^{\nu\sigma}\right) \right\} \,,
 \label{hder}
 \end{equation}
we obtain
%
%
 \begin{equation}
(-g)\left\{ T^{\mu\nu}_{\rm m} - \frac{4}{\kappa^2}\left( R^{\mu\nu}- \frac{1}{2} \, g^{\mu\nu} R + \Lambda \, g^{\mu\nu} \right) + {\cal T}^{\mu\nu} \right\} = 
 (-g) \, T^{\mu\nu} - \frac{2}{\kappa^2} \frac{\partial^2}{\partial x^\sigma \partial x^\lambda} \left\{ (-g) \left( g^{\mu\nu} g^{\lambda\sigma} - g^{\mu\lambda} g^{\nu\sigma}\right) \right\}   \,.
\label{def2}
\end{equation}
Choosing $f(g)=2\sqrt{-g}$ from Eqs.~(\ref{trans}), (\ref{EEOM}) and (\ref{def2}) we obtain
\begin{equation}
\int {\cal D}h {\cal D}\psi {\cal D} \bar c {\cal D} c 
\biggl( 
  \sqrt{-g} \,  g^{\mu\nu}(x)  \delta^{(4)}(0) 
 - \frac{i}{\hbar}\,\frac{2}{\kappa^2} \frac{\partial^2}{\partial x^\sigma \partial x^\lambda} \left\{ (-g) \left( g^{\mu\nu} g^{\lambda\sigma} - g^{\mu\lambda} g^{\nu\sigma}\right) \right\}  
 + \frac{i}{\hbar}\, (-g) \, T^{\mu\nu} (x) \biggr) e^{\frac{i}{\hbar} S(g,\psi,\bar c,c)} =0\,.
\label{cond1}
\end{equation}
Using the expansion around the Minkowski background and employing
dimensional regularization, the first term in Eq.~(\ref{cond1})
involving $\delta(0)$ vanishes. In the second term the derivatives can be taken out of the path integral. Calculating the integral 
by perturbative expansion around the Minkowski background one obtains result which does not depend on the spacetime coordinates due to translational invariance of the Minkowski background. By acting with derivatives on this constant expression one gets vanishing result.
The remaining third term corresponds to the vacuum energy, which hence turns out to vanish in all orders of perturbation theory.  

To summarize, the cosmological constant term, viewed  as a parameter
of the effective Lagrangian, is known to be uniquely fixed from the
consistency condition of the theory in Minkowski background
\cite{Burns:2014bva}.    
On the other hand, the condition of vanishing of the vacuum energy
also fixes the cosmological constant parameter uniquely. In this
paper, we have shown that the resulting two values of the cosmological
constant coincide to all orders in perturbation theory.
Thus, one can conclude that by demanding the vanishing of the vacuum energy within perturbation theory one is uniquely lead to a consistent EFT in the Minkowski background. 
The resulting
effective action, expressed
in terms of local operators, is free of the effective cosmological constant term. 

%


\acknowledgments

 We thank O.~Sakhelashvili for useful comments on the manuscript.
This work was supported in part 
by the MKW NRW under the funding code NW21-024-A, by the Georgian
Shota Rustaveli National 
Science Foundation (Grant No. FR-23-856), 
by the European Research Council (ERC) under the European Union's Horizon 2020 research and innovation programme (grant agreement No. 885150),
by CAS through a President's International Fellowship Initiative (PIFI)
(Grant No. 2025PD0022),  and by the EU Horizon 2020 research and
innovation programme (STRONG-2020, grant agreement No. 824093).



\end{document}